\documentclass[conference]{IEEEtran}
\IEEEoverridecommandlockouts
\IEEEpubid{\makebox[0.5\columnwidth]{978-1-7281-1867-3/19/\$31.00~\copyright2019 IEEE \hfill} \hspace{\columnsep}\makebox[\columnwidth]{ }}

\usepackage{cite}
\usepackage{amsmath,amssymb,amsfonts}
\usepackage{algorithmic}
\usepackage{graphicx}
\usepackage{textcomp}
\usepackage{tabularx}
\usepackage{booktabs}
\usepackage{hyperref}
\usepackage{graphics}

\usepackage[dvipsnames]{xcolor}
\def\BibTeX{{\rm B\kern-.05em{\sc i\kern-.025em b}\kern-.08em
    T\kern-.1667em\lower.7ex\hbox{E}\kern-.125emX}}
\definecolor{purple}{rgb}{0.29, 0.0, 0.51}
\definecolor{electricviolet}{rgb}{0.56, 0.0, 1.0}

\begin{document}
\title{Temporary title\\
\title{Supporting supervised learning in \\ fungal Biosynthetic Gene Cluster discovery: \\ new benchmark datasets } 
\thanks{NSERC}
}

\author{\IEEEauthorblockN{
Hayda Almeida$^{1,2}$}
 \IEEEauthorblockA{
 }
\and
\IEEEauthorblockN{
Adrian Tsang$^{2}$} \vspace{-0.25cm}
 \IEEEauthorblockA{
 }
\textit{$^{1}$University of Quebec in Montreal, Montreal, Canada} \\  
\textit{$^{2}$Concordia University, Montreal, Canada}\\
Corresponding author: diallo.abdoulaye@uqam.ca \\
\and
\IEEEauthorblockN{
Abdoulaye Banir\'e Diallo$^{1}$}
 \IEEEauthorblockA{
 }
}
\maketitle

\begin{abstract}
Fungal Biosynthetic Gene Clusters (BGCs) of secondary metabolites are clusters of genes capable of producing natural products, compounds that play an important role in the production of a wide variety of bioactive compounds, including antibiotics and pharmaceuticals.
Identifying BGCs can lead to the discovery of novel natural products to benefit human health.
Previous work has been focused on developing automatic tools to support BGC discovery in plants, fungi, and bacteria.
Data-driven methods, as well as probabilistic and supervised learning methods have been explored in  identifying BGCs.
Most methods applied to identify fungal BGCs were data-driven and presented limited scope.
Supervised learning methods have been shown to perform well at identifying BGCs in bacteria, and could be well suited to perform the same task in fungi.
But labeled data instances are needed to perform supervised learning.
Openly accessible BGC databases contain only a very small portion of 
 previously curated fungal BGCs.
Making new fungal BGC datasets available could motivate the development of supervised learning methods for fungal BGCs and potentially improve prediction performance compared to data-driven methods.
In this work we propose new publicly available fungal BGC datasets to support the BGC discovery task using supervised learning.
These datasets are prepared to perform binary classification and predict candidate BGC regions in fungal genomes.
In addition we analyse the performance of a well supported supervised learning tool developed to predict BGCs.
\end{abstract}


\section{Introduction}
\label{sec:intro}
Natural products (NPs) are specialized bioactive compounds primarily produced by plants, fungi and bacteria. 
NPs are a vital source for 
drugs: from anti-cancer, anti-virus, and cholesterol-lowering medications to antibiotics, and immunosuppressants~\cite{chaudhary2013insight,medema2015computational,chavali2017bioinformatics}. 
Unlike those in plants, genes involved in the biosynthesis of many NPs in bacteria and fungi are co-localized in the genome of organisms and usually organized as clusters of genes~\cite{osbourn2010secondary}.
Gene clusters capable of producing NPs are known as Biosynthetic Gene Clusters (BGC).

The task of identifying new BGCs could potentially lead to the discovery of novel NPs to benefit human health. 
However this task involves complex and costly processes, as well as the analysis of large amounts of biological data.
Development of automatic tools that can support the identification of BGCs is therefore highly relevant.
Various approaches have been used to develop such tools, such as data-driven methods, probabilistic methods, and supervised learning methods.
In supervised learning the BGC discovery task can be represented as binary classification task. 
The goal in a binary classification task is to classify data instances as belonging to one out of two different categories.
A binary classification BGC dataset would therefore be composed of positive and negative BGC instances.

Supervised learning has been previously used to predicting bacterial BGCs~\cite{agrawal2017rippminer,hannigan2019deep} and shown to perform well.
Supervised learning methods however are developed primarily based on annotated datasets, for which all instances are labeled as belonging to a specific class.
Unlike for bacteria, 
the number of known fungal BGC data previously validated by curators is 
rather limited. 
The Minimum Information about a Biosynthetic Gene cluster (MIBiG)~\cite{medema2015mibig}\footnote{\url{http://mibig.secondarymetabolites.org/}} repository is one of the largest freely available BGC databases. 

As an example of the disparity between 
known available BGC from bacteria versus fungi that has been annotated by curators,
MIBiG holds over 1,196 bacteria BGCs, while only 206 are fungal BGCs\footnote{As of July 2019.}.

Generating fungal BGC datasets for supervised learning approaches imposes a few challenges. 
For instance, negative samples are needed for binary classification, and they are not directly provided by BGC databases just as annotaded BGC data.
To be able to support a robust classification approach, fungal BGC datasets used as input should include a variety of organisms and BGC types to properly represent fungal genomic profiles.

The availability of fungal BGC datasets could leverage the development of new supervised learning approaches to tackle BGC discovery in fungi. 
This work presents new datasets prepared to tackle fungal BGC discovery as a binary classification task.
These datasets are constructed in such way that they include most variety of BGC types from different organisms, attempting to represent fungal genomic profiles to better suit the fungal BGC classification task. 
Finally we also analyse the usage of fungal BGC datasets with one of the state-of-the-art supervised learning methods developed for BGC discovery, DeepBGC~\cite{hannigan2019deep}.

\section{Previous Work}
In this section we present previous work on the availability of BGC data 
previously predicted or annotated by curators
that can support BGC discovery, and previous work conducted towards developing automatic approaches to identify fungal BGCs.
BGC databases and some of their characteristics are discussed in Section~\ref{subsec:databases}.
Previous work on predicting BGCs in fungi is presented in Section~\ref{subsec:fungi}.

\subsection{BGC Databases}
\label{subsec:databases}
Only a small number of open access BGC databases is currently available to support research on automatic tools to identify BGCs.
The majority of entries in these databases corresponds to bacteria data, while only a small portion are fungal BGCs.\footnote{Number of entries for databases are reported as of July 2019.}
MIBiG is a BGC repository in which curated entries are submitted by curators, and added to the database in a format compliant with the Minimum Information about any Sequence (MIxS) framework data standard.
It holds 206 fungi BGCs and 1,196 for bacteria.
Clustermine360~\cite{conway2012clustermine360} contains microbial polyketide synthases (PKS) and non-ribosomal peptide synthetases (NRPS) biosynthesis. 
It holds a total of 29 fungal BGCs, while over 900 are from bacteria.
Clustermine360 entries are curated and submitted by curators, enriched with information from the National Center for Biotechnology Information (NCBI)\footnote{https://www.ncbi.nlm.nih.gov/}, and analysed with the antiSMASH~\cite{blin2017antismash} tool.
The antiSMASH database~\cite{blin2016antismash} has 24,773 microbial BGCs predicted based on its homonymous tool.
Unlike its bacteria version, the fungal version of antiSMASH does not provide a database of fungal BGCs to the best of our knowledge.

The Integrated Microbial Genomes: Atlas of Biosynthetic Gene Clusters~\cite{hadjithomas2016imgabc} (IMG/ABC) database contains BGCs predicted using the ClusterFinder algorithm~\cite{cimermancic2014clusterfinder}.
IMG/ABC holds 127 fungal BGCs and 1,025 from bacteria.

These databases are not connected. 
Since it is likely that there are overlaps among the different databases, the number of unique fungal BGCs could be even smaller.
The small proportion of fungal BGCs across databases is an example of the challenges in developing
automatic tools to tackle BGC discovery in fungi.
This work proposes new publicly available datasets to be an input of supervised learning tools to predict fungal BGCs, based on MIBiG and orthologous genes.
The details on our datasets and their analysis are discussed in Section~\ref{sec:methodology}.

\subsection{BGC discovery in Fungi}
\label{subsec:fungi}
Significant effort has been put towards developing approaches to discover BGCs~\cite{medema2015computational,chavali2017bioinformatics}.
The majority of approaches focused on processing bacterial data, while some of them are specially focused on fungi.
Identifying BGCs remains a challenging task specially in fungal genomes, due to the diversity of clusters~\cite{kjaerbolling2018linking}.

Previous work on fungal BGC discovery made use mostly of data-driven methods, which are heavily based on the analysis of the input or output data and require fine parameter-tuning.
These methods required as input the genome sequence combined with transcription data~\cite{vesth2016fungeneclusters, umemura2013middas}, or gene functional annotations~\cite{wolf2016cassis}, as well as both nucleotide and amino acid sequences~\cite{takeda2014motif}. 
\cite{vesth2016fungeneclusters} and~\cite{umemura2013middas} focused on analysing similar gene expression levels, while~\cite{umemura2013middas} used virtual clusters.
\cite{vesth2016fungeneclusters} looked at motif co-occurrence in promoters around anchor genes, and~\cite{takeda2014motif} analysed homologous genes through a comparative genomics approach.

Such data-driven methods are less dependent on 
curated
BGC data, which 
are time consuming to obtain, but they all present limitations. 
~\cite{wolf2016cassis} requires gene functional annotations, which may not be available,
and~\cite{vesth2016fungeneclusters} 
relies heavily on manual curation of output to achieve the expected results.
A very limited BGC prediction scope is considered in ~\cite{khaldi2010smurf} and~\cite{takeda2014motif}.
Both approaches are developed based on biological sequences from a single species, and they also require fine parameter-tuning. 
Such limitations of data-driven methods can restrict their ability to generalize to new data, and as a consequence compromise the discovery of novel BGCs.

Likely due to the larger availability of 
curated
BGC data, probabilistic~\cite{cimermancic2014clusterfinder, skinnider2015prism, blin2017antismash} and machine learning approaches~\cite{agrawal2017rippminer,hannigan2019deep} have been more explored in bacteria compared to fungi, and shown to perform well.
Probabilistic and machine learning approaches could be beneficial for BGC discovery, since by nature they are more capable of generalizing given new data, and will likely perform better at identifying data patterns and discovering novel BGCs, when compared to data-driven methods.
In this study we also analyse the performance of a supervised learning approach developed to tackle BGC discovery using the fungal BGC datasets proposed by our work.
The details on our experimental setup are further discussed in Section~\ref{sec:methodology}.


\section{Methodology}
\label{sec:methodology}
Some of the challenges in generating fungal BGC datasets for binary classification are the need of negative instances, which are not directly provided in BGC databases; and accounting for a variety of organisms, BGC types, and also fungal genomic profiles.
The availability of new fungal BGC datasets however could potentially motivate the development of supervised learning approaches to tackle fungal BGC discovery.

In this work we propose new publicly available fungal BGC datasets to support supervised learning approaches tackling BGC discovery as a binary classification task.
We present here the methodology adopted to prepare fungal BGC datasets and their analysis using a supervised learning method, with the goal of analysing the method performance in fungal BGC data.

Details on our proposed fungal BGC datasets are presented in Section~\ref{subsec:datasets}.
Section~\ref{subsec:testdata} presents the test datasets with which we analysed the performance of classification models built on fungal BGC datasets.
In Section~\ref{subsec:classifmodels} we provide details on the parameters considered in our analysis based on a supervised learning method, as well as the classification models considered.

\subsection{Proposed Datasets}
\label{subsec:datasets}
Supervised learning was shown to perform well at BGC discovery in previous work that focused on handling bacteria data~\cite{agrawal2017rippminer,hannigan2019deep}.
Given that annotated data are needed to perform a supervised learning approach, we propose here fungal BGC datasets to support the development of this approach for fungi.

As mentioned in Section~\ref{sec:intro}, positive and negative instances are needed to perform fungal BGC discovery as a binary classification task using supervised learning.
To create our fungal BGC datasets, we extracted and filtered positive instances from the MIBiG~\cite{medema2015mibig} repository, previously presented in Section~\ref{subsec:databases}.
MIBiG has the highest number of unique fungal BGCs among the BGC databases previously presented.
Additionally, MIBiG BGCs were annotated and submitted by the research community, unlike BGCs in other databases that were automatically predicted.

From all MIBiG instances, we have selected only the fungal BGC subset, excluding BGCs belonging to \textit{Aspergillus niger} (\textit{A. niger}) to avoid overlaps during the test phase, resulting in a total of 200 positive instances.

We generated synthetic negative instances collecting and integrating orthologous genes from OrthoDB\footnote{\url{http://orthodb.org/}}~\cite{kriventseva2018orthodb}. 
Orthologs are homologous genes descendants from a single gene of a last common ancestor.
The OrthoDB database contains protein-coding genes that represent the last common ancestors given a specific phylogeny radiation of a species, and are therefore known 
 to retain ancestral function~\cite{kriventseva2018orthodb}.
Orthologs represent regions conserved across species. They can correspond to a  relevant negative instances for BGC discovery.
this is due to the fact that fungal BGCs are known to have opposite characteristics and show large genomic diversity even in otherwise closely-related or same genus species~\cite{kjaerbolling2018linking}.
Genes belonging to fungal BGCs have been previously referred to as ``species-specific"~\cite{vesth2018investigation}, unlike orthologs.

Orthologous genes have been previously used to discover BGCs in fungi.
In~\cite{takeda2014motif}, the authors presented an alignment-based approach focused on identifying syntenic block regions, which are more likely to contain orthologs and less likely to contain BGCs.
Non-syntenic blocks were then used to search for candidate BGCs and to better define candidate cluster boundaries. 
The approach in~\cite{takeda2014motif} was explored in small set of 10 filamentous fungi. 
The results showed good performance, predicting correctly 21 out of 24 fungal BGCs.

In this study we selected the fungal OrthoDB subset to construct the synthetic negative BGC instances.
The OrthoDB fungal subset contains a total of 5,083,652 non-redundant orthologs. 
To avoid potential overlaps, we performed a BLAST analysis between the fungal subsets of both OrthoDB and MIBiG. We discarded 11,000 ortholog matches found using the BLAST parameter $evalue$ (expected value) set to $1e-60$. 

\begin{figure}[!htbp]
    \centering
\includegraphics[width=0.47\textwidth]{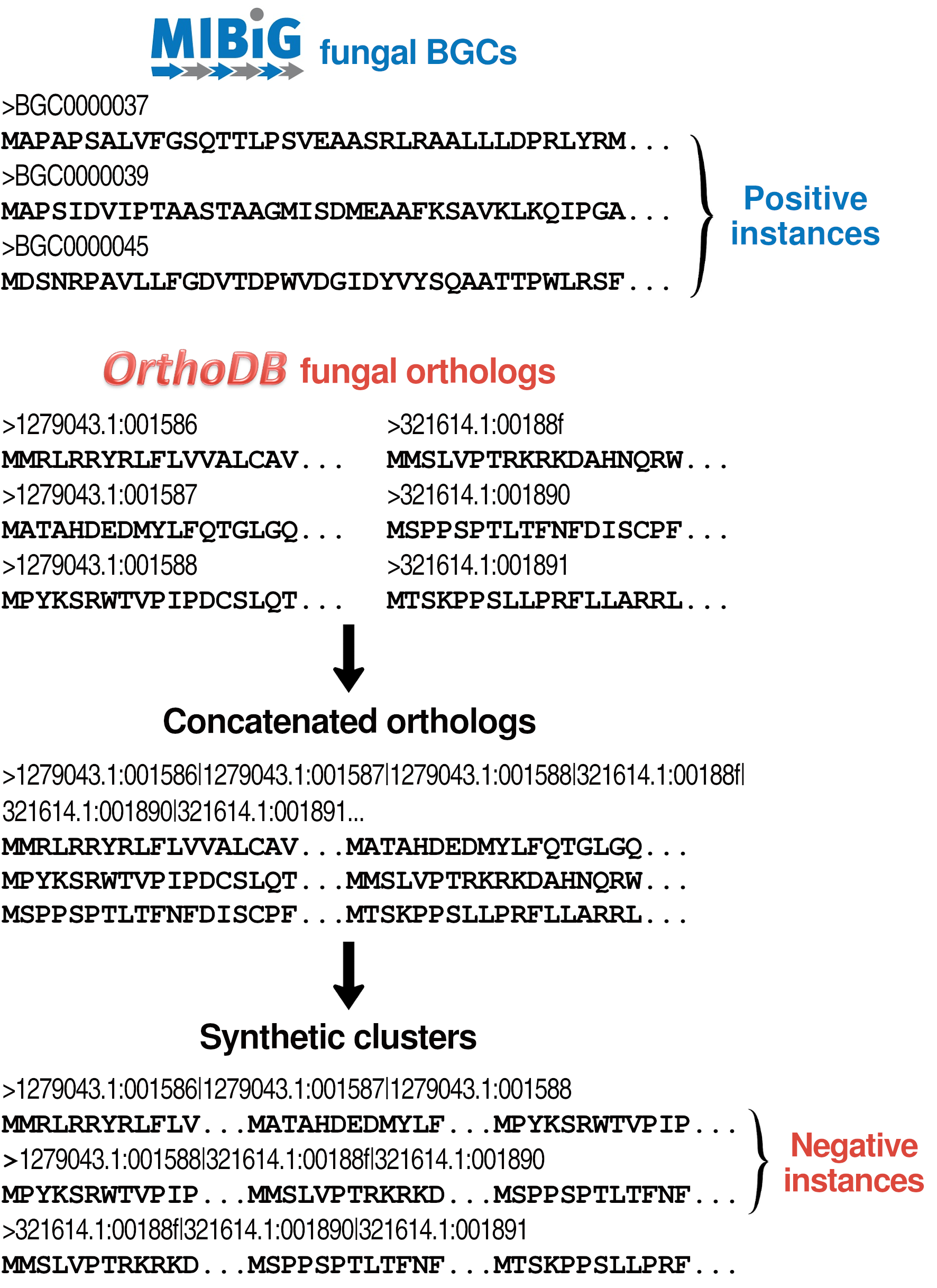}
    \caption{Example of positive instances and the process to generate synthetic negative instances from orthologs}
    \label{fig:traindata}
\end{figure}

To generate synthetic negative instances, we then concatenated the amino acid sequence of fungal orthologs using a fixed length of 7,000 amino acids to create synthetic gene clusters.
The 7,000 amino acid length is chosen since it corresponds to the average length of fungal BGC amino acid sequences in MIBiG.
Figure~\ref{fig:traindata} shows an example of positive instances in our datasets and negative instances being generated from OrthoDB orthologs.
After processing OrthoDB fungal orthologs a total of 693,195 synthetic negative clusters were generated.

The MIBiG fungal subset and the pool of OrthoDB synthetic negative clusters were then considered to generate fungal BGC datasets with different distributions of positive and negative instances.
Among the MIBiG fungal subset the annotated BGC regions corresponded in average to $\approx$1\% of the total genome length of an organism, which provides a hint on the imbalance in class distribution that can be seen in a real test case scenario.
Due to the natural imbalance of BGC regions versus non-BGC regions in a genome, we are interested in analysing the performance of a supervised learning approach based on datasets with various distributions of positive and negative instances.
To analyse this aspect, we generated fungal BGC datasets with varying distributions by increasing the number of synthetic negative instances randomly selected from the OrthoDB synthetic negative clusters pool.
Table~\ref{tab:datasets} shows the positive vs. negative distributions in each dataset.

\begin{table}[!htbp]
\centering
\caption{Distribution of instances across fungal BGC datasets }
\label{tab:datasets}
\begin{tabular}{|c|rr|rr|} \hline
    & \multicolumn{2}{c|}{\bf Train} & \multicolumn{2}{c|}{\bf Validation} \\
\bf Dataset & \bf Pos   & \bf Neg   & \bf Pos   & \bf Neg \\ \hline
50\%-50\%   & 160       & 160       & 40        & 40 \\
40\%-60\%   & 160       & 240       & 40        & 60 \\
30\%-70\%   & 160       & 373       & 40        & 93 \\
20\%-80\%   & 160       & 640       & 40        & 160 \\
10\%-90\%   & 160       & 1,440     & 40        & 360 \\
05\%-95\%   & 160       & 3,040     & 40        & 760 \\
01\%-99\%   & 160       & 15,840    & 40        & 3,960 \\ \hline
    \end{tabular}
\end{table}

To generate classification models based on a supervised learning method, we extracted Pfam~\cite{el2018pfam}\footnote{\url{http://pfam.xfam.org}} IDs from the positive and negative instances.
All datasets were converted into \texttt{pfamtsv} format~\cite{hannigan2019deep}, which is required as input in the supervised learning approach applied in this work.
For each dataset, 80\% were randomly selected for the training phase, while 20\% were held out for the validation phase, as shown in Table~\ref{tab:datasets}.

\subsection{Test Datasets}
\label{subsec:testdata}
To analyse the performance of classification models built based on fungal BGC datasets, we selected a fungal genome from the \textit{Aspergillus} genus to represent a real test case scenario. 
\textit{Aspergillus} is the most frequent genus among fungal species in MIBiG, together with \textit{Penicillium}.
For this evaluation we focused specifically on the \textit{A. niger} species.
\textit{A. niger} is a genome of interest due to its biological diversity and major relevance to industrial processes~\cite{de2017comparative}.
In~\cite{inglis2013comprehensive} the authors present manual annotation of BGCs in \textit{Aspergilli}, among which a total of 79 BGCs are found in \textit{A. niger}.

To generate candidate clusters for the test phase, we collected a manually curated \textit{A. niger} genome sequence made publicly available through the Genozymes project\footnote{https://gb.fungalgenomics.ca/portal/}.
We generated test candidate clusters by considering a sliding window of 30,000 nucleotides in the \textit{A. niger} genome. 
The 30,000 sliding window length is defined based on the average length of the nucleotide sequence of MIBiG fungal BGCs.
A similar approach was previously applied in fungal BGC discovery to generate virtual clusters~\cite{umemura2013middas}.

\begin{figure}[!htbp]
    \centering
    \includegraphics[width=0.45\textwidth]{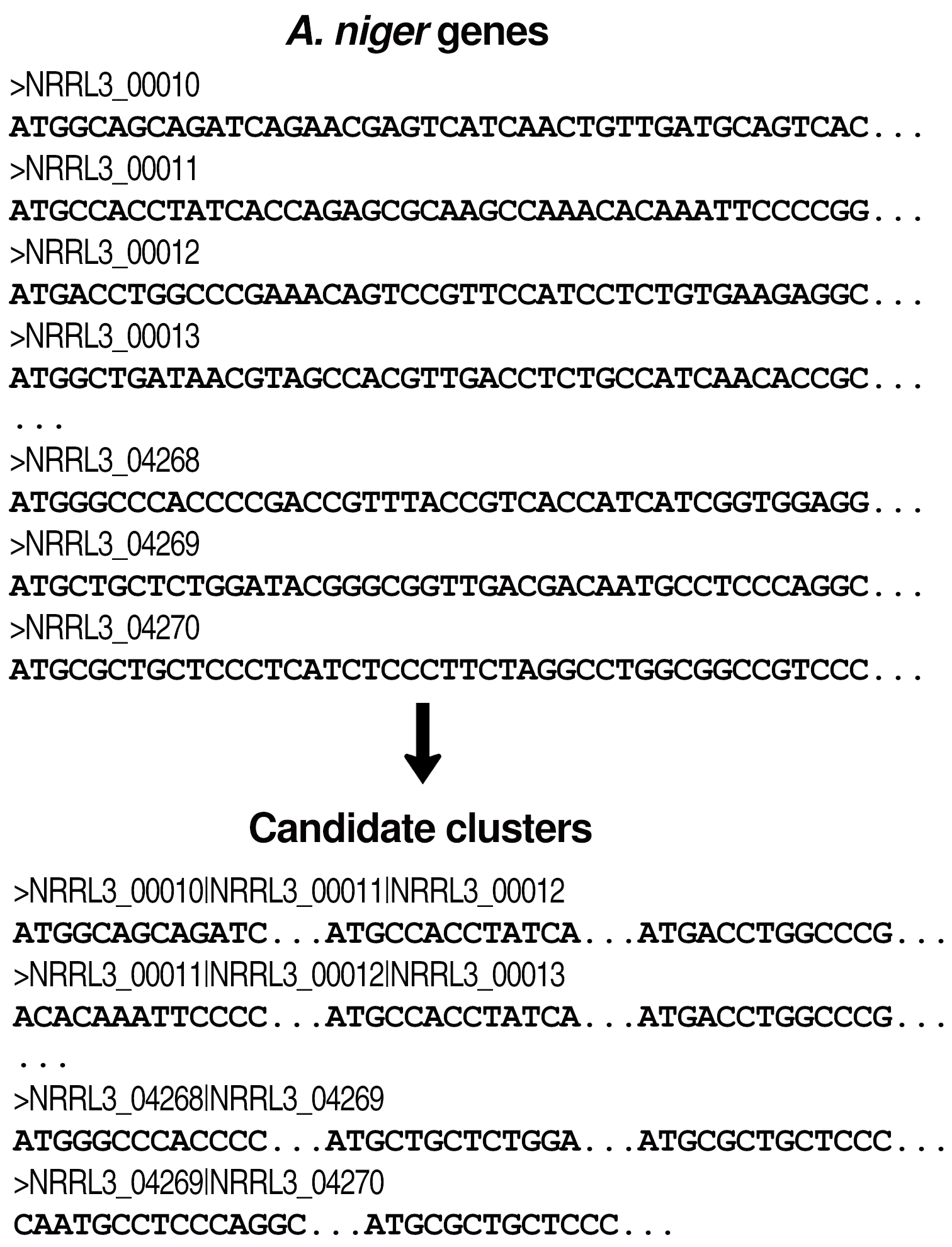}
    \caption{Example of \textit{A. niger} candidate clusters generated for test phase}
    \label{fig:testdata}
\end{figure}

The 30,000 sliding window was shifted along the genome using either a 50\% or a 30\% overlap.
The overlaps in a sliding window mean that each test candidate cluster will contain the last 15,000 nucleotides (if a 50\% overlap) or the last 9,000 nucleotides (if a 30\% overlap) of the immediate previous candidate cluster.
With the strategy of generating candidate clusters using overlaps, we are more likely to cover regions in between two or more genes.
Figure~\ref{fig:testdata} shows an example of candidate clusters being generated from \textit{A. niger} genes using overlaps.
The test datasets based on a 50\% overlap contains a total of 1,184 candidate clusters, while the test datasets based on a 30\% overlap contains a total of 846 candidate clusters.

\subsection{Classification Models}
\label{subsec:classifmodels}
In this section we describe the methods applied to analyse the performance of a supervised learning approach using the fungal BGC datasets presented in Section~\ref{subsec:datasets} and the test data presented in Section~\ref{subsec:testdata}.
To generate classification models with our fungal BGC datases, we utilized the DeepBGC system~\cite{hannigan2019deep}.
DeepBGC executable, source code and other resources are openly available\footnote{https://github.com/Merck/deepbgc}.
Among these resources, there are pre-built BGC classification models and word2vec-based embeddings built using Pfam IDs, referred to as pfam2vec embeddings.
In~\cite{hannigan2019deep} the authors explained that pfam2vec embeddings were trained based in a skipgram architecture with 100 dimensions and over 15,686 unique Pfam IDs.
DeepBGC classification is based on a Bidirectional Long Short Term Memory (BiLSTM) neural network, for which the input are pfam2vec embeddings. 
In~\cite{hannigan2019deep} DeepBGC hyperparameters are described as a BiLSTM layer size of 128, dropout of 0.2, sigmoid activation, batch size of 64, 256 timestamps over 328 epochs, using Adam optimizer at a learning rate of 1e-4, with weighted binary cross-entropy loss.
To generate classification models using fungal BGC datasets on the DeepBGC system we adopted the same hyperparameters described in~\cite{hannigan2019deep}, as well as the pfam2vec embeddings as input for training.
For each fungal BGC dataset, we have generated a different classification model using DeepBGC.
Fungal BGC models are named by their positive instance percentage:
\begin{itemize}
    \item \texttt{pos50}  (50\%-50\%)
    \item \texttt{pos40}  (40\%-60\%)
    \item \texttt{pos30}  (30\%-70\%)
    \item \texttt{pos20}  (20\%-80\%)
    \item \texttt{pos10}  (10\%-90\%)
    \item \texttt{pos05}  (05\%-95\%)
    \item \texttt{pos01}  (01\%-99\%)
\end{itemize}

To complement our analysis, we also analysed the performance of our test datasets using the four bacteria-based models made available at the DeepBGC repository: 
\begin{itemize}
    \item \texttt{deepbgc} 
    \item \texttt{cf\_o} (clusterfinder\_original)
    \item \texttt{cf\_r} (clusterfinder\_retrained)
    \item \texttt{cf\_g} (clusterfinder\_geneborder) 
\end{itemize}

According to the models description at the DeepBGC releases page\footnote{https://github.com/Merck/deepbgc/releases} and~\cite{hannigan2019deep}, the \texttt{deepbgc} model is based on the BiLSTM DeepBGC architecture and trained on a MIBiG dataset.
The other models are built based on  ClusterFinder~\cite{cimermancic2014clusterfinder}, which is a Hidden Markov Model (HMM). \texttt{cf\_o} is a ClusterFinder HMM using original parameters; \texttt{cf\_r} is also a ClusterFinder HMM but trained on a MIBiG dataset; and \texttt{cf\_g} is a ClusterFinder HMM that switches stages only on gene borders, and trained on a MIBiG dataset.

\section{Results and discussion}
\label{sec:results}
We present here statistics and further details on the publicly available fungal BGC datasets proposed in this study.
We also present results of validation and test phase obtained with classification models based on fungal BGC datasets and built using DeepBGC.
Section~\ref{subsec:datastats} has further information and statistics on the fungal BGC datasets proposed in our work.
In Section~\ref{subsec:valperf} we present results obtained at validation of training DeepBGC using the models \texttt{pos50}, \texttt{pos40}, \texttt{pos30}, \texttt{pos20}, \texttt{pos10}, \texttt{pos05}, and \texttt{pos01}.
In Section~\ref{subsec:testperf} we present results obtained at test phase. 
For the sake of comparison, we also report the results on test data using BGC classification models provided by DeepBGC and built based on bacteria data, as listed in Section~\ref{subsec:classifmodels}.
All performance metrics are reported on the positive class only.

\subsection{Fungal BGC datasets}
\label{subsec:datastats}

The fungal BGC datasets proposed in this work are composed of positive and negative instances, as mentioned in Section~\ref{subsec:datasets}.
These datasets are suitable for performing binary classification to predict fungal BGCs, and are made publicly available at \url{https://github.com/bioinfoUQAM/fungalbgcdata}.
The availability of such resource can potentially motivate the development of supervised learning approaches to tackle BGC discovery in fungi.

Positive instances in our datasets represent fungal BGCs from 52 different fungal genera.
The variety of fungal genus is relevant to provide a large representation of BGC occurrence through different organisms.
Additionally, the positive instances contain samples of over 10 different BGC types.
Table~\ref{tab:positive} shows the different BGC types and a summary of fungal genera in our datasets. 
As the table shows, the most common BGC type is Polyketide synthase (PKS), followed by Non-ribosomal peptide synthase (NRP) and Terpene synthase (TC).
The presence of different fungal genus and BGC types in the datasets are important for representing a wide variety of BGC occurrences, and therefore contribute to building more robust supervised learning approaches.  

\begin{table}[!htbp]
\centering
\begin{minipage}{.33\linewidth}
\scalebox{0.9}{
\begin{tabular}{lr}
\toprule
\bf BGC types        & \# \\ \midrule
Alkaloid        & 3 \\
Alkaloid/NRP    & 3 \\
Alkaloid/TC     & 1 \\
Alkaloid/NRP/TC & 1 \\
NRP             & 41 \\
NRP/PKS         & 19 \\
PKS             & 90 \\
PKS/TC          & 5 \\
RiPP            & 3 \\
Saccharide      & 1 \\
TC              & 23 \\
Other           & 10 \\ \midrule
Total           & \bf 200 \\ 
\end{tabular}}
\end{minipage}
\begin{minipage}{.62\linewidth}
\scalebox{0.9}{
\begin{tabular}{lr lr}
\toprule
\bf BGC fungi genus & \# & \bf BGC fungi genus  & \# \\ \midrule
Acremonium      & 1     & Metacordyceps     & 1  \\
Alternaria      & 5     & Metarhizium       & 1  \\
Armillaria      & 1     & Monascus          & 3 \\
Aspergillus     & 9    & Mycosphaerella    & 1  \\
Aureobasidium   & 1     & Myrothecium       & 1  \\
Beauveria       & 1     & Neosartorya       & 1  \\
Bipolaris       & 3     & Neotyphodium      & 2 \\
Botrytis        & 1     & Nodulisporium     & 1  \\
Byssochlamys    & 1     & Paecilomyces      & 1  \\
Cercospora      & 1     & Parastagonospora  & 1  \\
Chaetomium      & 2     & Penicillium       & 13 \\
Cladonia        & 2     & Pestalotiopsis    & 1  \\
Claviceps       & 2     & Phoma             & 2 \\
Diaporthe       & 1     & Phomopsis         & 1  \\
Elsinoe         & 1     & Purpureocillium   & 1  \\
Epichloe        & 2     & Sarocladium       & 1  \\
Fusarium        & 8     & Shiraia           & 1  \\
Glarea          & 1     & Sordaria          & 1  \\
Glycomyces      & 1     &  Sphaceloma       & 1  \\
Hypholoma       & 1     & Stachybotrys      & 1  \\
Hypomyces       & 1     & Starmerella       & 1  \\
Isaria          & 1     & Talaromyces       & 3 \\
Lasiodiplodia   & 1     & Tapinella         & 1  \\
Lecanicillium   & 1     & Tolypocladium     & 2 \\
Leptosphaeria   & 1     & Trichophyton      & 1  \\
Malbranchea     & 1     & Ustilago          & 1  \\ 
\vspace{0.005cm}
\end{tabular}}
\end{minipage}
\caption{Fungal genera and BGC types in positive instances of datasets}
\label{tab:positive}
\end{table}

Negative instances in our datasets represent synthetic gene clusters composed of fungal orthologs.
By using fungal orthologs as source for the negative instances, we can generate synthetic gene clusters that  depict the genomic profile of fungi.
A total of 549 fungal species are present in orthologs composing our negative instances.
The main fungal groups to which the orthologs belong to are shown in Table~\ref{tab:negative}, according to their taxonomy level.
In this table we show the number of species clustered under different taxonomy levels (genus, family, order, or class), and the corresponding total of non-redundant orthologous genes for each group.
\begin{table}[!htbp]
\centering
\scalebox{0.9}{
\begin{tabular}{llrr}
\toprule
\bf Group   & \bf Taxonomy & \bf \# Species & \bf \# Genes \\
            & \bf level & & \\ 
\midrule
Aspergillus         & Genus     & 30    & 309,629 \\
Cryptococcus        & Genus     & 7     & 44,028 \\ 
Exophiala           & Genus     & 7     & 67,291 \\ 
Metarhizium	        & Genus     & 5     & 45,563 \\ 
Penicilium          & Genus     & 21    & 208,580 \\
Phytophthora	    & Genus	    & 6     & 89,378 \\
Hypocreaceae        & Family    & 7     & 66,815 \\ 
Pleosporaceae       & Family    & 9     & 94, 817 \\
Polyporaceae        & Family    & 6     & 61,584 \\ 
Saprolegniaceae	    & Family	& 6     & 81,114 \\
Trichocomaceae      & Family    & 6     & 52,941 \\ 
Agaricales	        & Order     & 25    & 293,149 \\
Eurotiales          & Order     & 60    & 608,401 \\
Helotiales	        & Order     & 14    & 162,251 \\
Hypocreales         & Order     & 50    & 512,282 \\
Mucorales           & Order     & 15    & 164,081 \\
Polyporales	        & Order     & 17    & 169,368 \\
Sordariales	        & Order     & 8     & 66,549 \\ 
Agaricomycetes      & Class     & 77    & 912,187 \\ 
Eurotiomycetes	    & Class     & 103   & 1,002,099 \\
Microbotryomycetes	& Class     & 9     & 59,326 \\
Pucciniomycetes	    & Class     & 6     & 64,018 \\
Saccharomycetes	    & Class     & 76    & 390,808  \\ 
Tremellomycetes	    & Class     & 18    & 121,702 \\
Ustilaginomycetes   & Class     & 9     & 55,465 \\ 
\vspace{0.005cm}
\end{tabular}}
\caption{Main fungal groups present in negative instances of datasets}
\label{tab:negative}
\end{table}

The 52 fungal genera in positive instances together with the 549 fungal species in negative instance orthologs contribute to represent the genomic diversity in fungi, and therefore support the development of more robust classification models.  

\subsection{Validation performance}
\label{subsec:valperf}
Table~\ref{tab:valresults} shows validation metrics obtained with fungal BGC datasets.
During training phase, all models using fungal BGC datasets had early stopping before completing the total 328 epochs.
This could be a sign that the models were overfitting, a possible consequence due to the size of the datasets and the imbalanced distribution between the two classes.

The best performing model \texttt{pos50} is the one with the most balanced distribution of positive and negative instances. 
It yield Precision (P) of 0.598, Recall (R) of 0.995, and F-measure (F) of 0.747.
Models \texttt{pos10}, \texttt{pos05}, and \texttt{pos01}, the ones with the most imbalanced distributions, had the lowest validation loss but also the lowest P, R and F.

\begin{table}[!htbp]
\begin{center}
\caption{Validation performance using models built on proposed datasets}
\label{tab:valresults}
\begin{tabular}{|c|r r r r r |} \hline
\bf Model    & \bf Epochs & \bf Loss    & \bf P     & \bf R     & \bf F   \\ \hline
\texttt{pos50}  & 91         & 0.683       & 0.598     & 0.995     & 0.747  \\
\texttt{pos40}  & 52         & 0.719       & 0.407     & 1         & 0.578   \\
\texttt{pos30}  & 108        & 0.667       & 0.536     & 0.743     & 0.623   \\ 
\texttt{pos20}  & 97         & 0.758       & 0.230     & 0.991     & 0.373  \\
\texttt{pos10}  & 70         & 0.389       & 0         & 0         & 0 \\    
\texttt{pos05}  & 73         & 0.240      & 0         & 0         & 0 \\
\texttt{pos01}  & 57         & 0.062      & 0         & 0         & 0 \\
\hline
\end{tabular}
\end{center}
\end{table}


\subsection{Test performance}
\label{subsec:testperf}
The test phase show how the models would perform in a real case scenario, when a complete genome is being processed to predict candidate BGC regions.
The dataset inputted in the test phase is composed of candidate clusters from the \textit{A. niger} genome sequence, as described in Section~\ref{subsec:testdata}.
The performance on the test data is presented in two ways: gene metrics and cluster metrics.
Gene metrics show P, R, and F for genes that belong to 
known
BGCs.
Cluster metrics show P, R, and F for 
known
BGCs where a minimum of one cluster gene must be correctly classified for the cluster to be predicted as positive.
Tables~\ref{tab:testresult50} and ~\ref{tab:testresult30} show the results on \textit{A. niger} test datasets, with overlaps of respectively 50\% and 30\%.
These results were obtained using classification models built with the fungal BGC datasets described in Section~\ref{subsec:datasets}.

\begin{table}[!htbp]
\begin{center}
\caption{Performance for \textit{A. niger} test data using models  built on fungal BGC datasets using 50\% overlap}
\label{tab:testresult50}
\begin{tabular}{|c|r r r|r r r|} \hline
        & \multicolumn{3}{c|}{\bf Gene metrics} & \multicolumn{3}{c|}{\bf Cluster metrics} \\ 
\bf Model    & \bf P & \bf R & \bf F & \bf P & \bf R & \bf F \\ \hline
\texttt{pos50} & 0.049 & 1.0 & 0.094 & 0.072 & 0.988 & 0.134 \\
\texttt{pos40} & 0.048 & 0.962 & 0.091 & 0.073 & 0.988 & 0.136 \\
\texttt{pos30} & 0.044 & 0.867 & 0.083 & 0.073 & 0.977 & 0.136 \\
\texttt{pos20} & 0.039 & 0.694 & 0.074 & 0.079 & 0.93 & 0.146 \\
\texttt{pos10} & 0 & 0 & 0 & 0 & 0 & 0 \\
\texttt{pos05} & 0 & 0 & 0 & 0 & 0 & 0 \\
\texttt{pos01} & 0 & 0 & 0 & 0 & 0 & 0 \\  \hline
\end{tabular}
\end{center}
\end{table}

\begin{table}[!htbp]
\begin{center}
\caption{Performance for \textit {A. niger} test data using models  built on fungal BGC datasets using 30\% overlap}
\label{tab:testresult30}
\begin{tabular}{|c|r r r|r r r|} \hline
        & \multicolumn{3}{c|}{\bf Gene metrics} & \multicolumn{3}{c|}{\bf Cluster metrics} \\ 
\bf Model       & \bf P & \bf R & \bf F & \bf P & \bf R & \bf F \\ \hline
\texttt{pos50}  & 0.05  & 1.0   & 0.096 & 0.1   & 0.988 & 0.182 \\
\texttt{pos40}  & 0.048 & 0.951 & 0.092 & 0.099 & 0.953 & 0.179 \\
\texttt{pos30}  & 0.045 & 0.865 & 0.085 & 0.1   & 0.942 & 0.18 \\
\texttt{pos20}  & 0.039 & 0.669 & 0.073 & 0.105 & 0.884 & 0.188 \\
\texttt{pos10}  & 0     & 0     & 0     & 0     & 0     & 0 \\
\texttt{pos05}  & 0     & 0     & 0     & 0     & 0     & 0 \\
\texttt{pos01}  & 0     & 0     & 0     & 0     & 0     & 0 \\  \hline
\end{tabular}
\end{center}
\end{table}

Results in the test phase show an important decrease in performance compared to the validation phase metrics.
However the behaviors observed at the validation step also appear in test.
Similarly to the validation phase, the more imbalanced models \texttt{pos10, pos05, pos01} did not predict any candidate cluster as positive.
This behavior happened with both test datasets of 50\% or 30\% overlap, and it could indicate that the model is sensitive to an imbalanced distribution of classes.

Also similarly to the validation phase the more balanced models \texttt{pos50, pos40, pos30, pos20} tended to predict most of candidate clusters as positives, leading to high recall but very low precision.
Table~\ref{tab:testresult30} shows slightly better performance for P, R, and F compared to table~\ref{tab:testresult50}.
This behavior could indicate that using a 30\% overlap in the test data is better suited for the task.

Following the results obtained with models based on fungal BGC datasets, we would like to also analyse the  performance of DeepBGC models built using bacteria data on \textit {A. niger} test datasets.
Tables~\ref{tab:testresult50deep} and~\ref{tab:testresult30deep} show the results obtained on \textit{A. niger} data with respectively 50\% and 30\% overlap using DeepBGC bacteria models.

\begin{table}[!htbp]
\begin{center}
\caption{Performance for \textit{A. niger} test data  with 50\% overlap using models provided by DeepBGC}
\label{tab:testresult50deep}
\begin{tabular}{|l|r r r|r r r|} \hline
       & \multicolumn{3}{c|}{\bf Gene metrics} & \multicolumn{3}{c|}{\bf Cluster metrics} \\ 
\bf Model           & \bf P     & \bf R     & \bf F     & \bf P     & \bf R     & \bf F \\ \hline
\texttt{deepbgc}    &  0.074    & 0.972     & 0.138     & 0.114     & 0.988     & 0.205  \\
\texttt{cf\_o}      &  0.05     & 1.0       & 0.096     & 0.074     & 0.988     & 0.138  \\
\texttt{cf\_r}      &  0.056    & 0.997     & 0.106     & 0.083     & 0.988     & 0.153  \\
\texttt{cf\_g}      & 0.06      & 0.989     & 0.113     & 0.09      & 0.988     & 0.166  \\ \hline
\end{tabular}
\end{center}
\end{table}

\begin{table}[!htbp]
\begin{center}
\caption{Performance for \textit{A. niger} test data  with 30\% overlap using models provided by DeepBGC}
\label{tab:testresult30deep}
\begin{tabular}{|l|r r r|r r r|} \hline
       & \multicolumn{3}{c|}{\bf Gene metrics} & \multicolumn{3}{c|}{\bf Cluster metrics} \\ 
\bf Model           & \bf P     & \bf R     & \bf F     & \bf P     & \bf R     & \bf F \\ \hline
\texttt{deepbgc}    & 0.074 & 0.954 & 0.138 & 0.159 & 0.988 & 0.273   \\
\texttt{cf\_o}      & 0.051 & 0.984 & 0.096 & 0.103 & 0.988 & 0.187  \\
\texttt{cf\_r}      & 0.058 & 0.994 & 0.109 & 0.118 & 0.988 & 0.211  \\
\texttt{cf\_g}      & 0.061 & 0.992 & 0.116 & 0.126 & 0.988 & 0.223 \\  \hline
\end{tabular}
\end{center}
\end{table}

Among all DeepBGC bacteria models, \texttt{deepbgc} performed best at both gene and cluster metrics, either using 30\% or 50\% overlap, with 0.273 F. The model \texttt{cf\_o} showed the lowest performance, with 0.138 F.
Models \texttt{cf\_r} and \texttt{cf\_g} showed in both cases better performance than \texttt{cf\_o}.
The results using DeepBGC trained models yield a similar tendency than that of the fungal BGC models: high recall but very low precision.
\\~\\
A loss in performance between validation and test results is evident, either when using fungal BGC based models or DeepBGC bacteria models.

As mentioned in Section~\ref{subsec:datasets}, fungal BGCs seem to show larger genomic diversity, which possibly makes it more complex to perform BGC discovery in fungi if compared to bacteria.
Therefore, performance is expected to be somehow affected by performing fungal BGC classification using bacteria-based models.

The dataset size at training time could also have had an impact on training \texttt{pos50, pos40, pos30, pos20, pos10, pos05} models, given DeepBGC classification approach.
As the authors in~\cite{angermueller2016deep} explained, the suitability of deep learning approaches varies according to the problem at hand; and in cases when available data is limited conventional approaches could be relevant and more advantageous.
As discussed in Section~\ref{subsec:datasets} 
the number of known fungal BGC data previously validated by curators is
rather limited,
which as a consequence will limit the size of fungal BGC datasets.
It is possible and worth investigating that different classification methods, apart from a BiLSTM neural network as adopted in DeepBGC, will be better suited for handling fungal BGC discovery.


\section{Conclusion}
\label{sec:conclusion}
NPs are bioactive compounds that play a vital role in the production of a large variety of drugs, and the discovery of novel NPs can potentially benefit human health.
Great effort has been put on identifying BGCs that are capable of producing NPs in plants, bacteria and fungi.
Identifying BGCs is a challenging task, specially in fungi given the clusters genomic diversity.

Previous work on identifying BGCs in bacteria have resulted in a large variety of approaches and annotated data available.
In fungi most previous approaches are based on data-driven methods and present a limited scope, such as covering only certain types of BGCs, or have been developed based on a single species data.
The availability of new fungal BGC datasets could potentially motivate the development of new methods to identify BGCs in fungi. 
One example is supervised learning, a method that have shown to perform well in bacteria data.

In this work, we present new fungal BGC datasets to leverage supervised learning in the fungal BGC discovery task. 
These datasets are made publicly available at \url{https://github.com/bioinfoUQAM/fungalbgcdata}.
The availability of such fungal BGC datasets can potentially motivate the development of binary classification approaches to tackle the BGC discovery task.
We have shown results obtained on these fungal BGC datasets using a supervised learning approach developed for bacteria BGCs. 
We also analysed the performance of bacteria-based classification models applied on a fungal genome.
The test performance on both fungal-based generated models or bacteria-based models was similar given precision (low) and recall (high) metrics using the same supervised learning method.
This points to an opportunity to explore different supervised learning approaches than the one adopted by the DeepBGC system, that might be more suitable to handle fungal BGC datasets.

\section*{Acknowledgment}
This work was supported by a fellowship of the Natural Sciences and Engineering Research Council of Canada (NSERC) to H.A., and NSERC Discovery Grant to A.B.D. and A.T. 

\bibliographystyle{ieeetr}
\bibliography{references}

\end{document}